\documentclass{article}

\usepackage{amssymb,amsmath}

\newcommand{\eqn}[1]{(\ref{#1})}
\newcommand{\be}{\begin{equation}}
\newcommand{\ee}{\end{equation}}
\newcommand{\beqa}{\begin{eqnarray}}
\newcommand{\eeqa}{\end{eqnarray}}

\title{On Reduced Time Evolution for Initially Correlated Pure States}
\author{P. Aniello \\ \footnotesize\it Dipartimento di Scienze Fisiche dell'Universit\`a `Federico II' e INFN, \\
\footnotesize\it Sezione di
Napoli, Complesso Universitario di Monte S. Angelo,\\ \footnotesize\it via Cintia,
80126 Naples, Italy \\ \footnotesize\it aniello@na.infn.it\\[2ex]
      A. Kossakowski \\ \footnotesize\it MECENAS, Universit\`a `Federico II',\\
      \footnotesize\it
via Mezzocannone 8, I-80134 Naples, Italy \\ 
\footnotesize\it kossak@fizyka.umk.pl\\[2ex]
G. Marmo \\ \footnotesize\it Dipartimento di Scienze Fisiche dell'Universit\`a  `Federico II' e INFN, \\ 
\footnotesize\it Sezione di
Napoli, Complesso Universitario di Monte S. Angelo, \\ \footnotesize\it via Cintia,
80126 Naples, Italy \\ \footnotesize\it marmo@na.infn.it\\[2ex]
F. Ventriglia \\ \footnotesize\it Dipartimento di Scienze Fisiche dell'Universit\`a `Federico II' e INFN,\\ 
\footnotesize\it Sezione di
Napoli, Complesso Universitario di Monte S. Angelo,\\ \footnotesize\it via Cintia,
80126 Naples, Italy \\ \footnotesize\it ventriglia@na.infn.it\\[2ex]
P. Vitale \\ \footnotesize\it Dipartimento di Scienze Fisiche dell'Universit\`a  `Federico II' e INFN, \\
\footnotesize\it 
Sezione di Napoli, Complesso Universitario di Monte S. Angelo,\\ \footnotesize\it
via Cintia,
80126 Naples, Italy \\ \footnotesize\it vitale@na.infn.it\\[2ex]
}
\date{}

\begin{document}

\maketitle

\begin{abstract}
A new method to deal with reduced dynamics of open systems  by means of the Schr\"odinger equation  
is presented. It allows one to consider the reduced time evolution for correlated and uncorrelated initial conditions.
\end{abstract}

\section{Introduction}

The study of time evolution of quantum open systems plays an
important role in quantum information. The interaction between the
system and the reservoir leads to phenomena of dissipation and
decoherence \cite{1,2}. These phenomena can be often modelled
through the standard technique of quantum Markov processes in
which the density matrix of an open system satisfies the Markovian
master equation \cite{3,4,5}. However, in the description of
complex quantum systems one encounters a complicated non-Markovian
behaviour [1,\,6\,--\,12]. 
Usually, non-Markovian
evolution is studied under the assumption that the initial
system-reservoir state is uncorrelated.

The reduced dynamics can be considered in the cases when the state of the
total system is mixed or pure. In the latter case, under the
assumption of the existence of some symmetries one can solve the
Schr\"odinger equation and deduce the reduced dynamics [13\,--\,21].
In the present paper a method to deal with
reduced dynamics by means of the Schr\"odinger equation is
presented. It allows one to study the reduced time evolution for
correlated and uncorrelated initial conditions, and also makes it
possible to consider models beyond the so called  rotating wave
approximation.

\section{A Class of Models}

We begin with the description of a class of models of open quantum systems
for which the reduced dynamics can be studied in the case when memory effects
as well as some initial correlations between the system and the reservoir are
taken into account. The method presented here is an extended version of \cite{22}.

Let us consider a $d$-level system coupled to the boson field. The associated Hilbert
space is $\mathbb{C}^d \otimes \mathcal{F}$ where $\mathcal{F}$ is the bosonic Fock space.
Let $|1\rangle,\ldots, |d\rangle$ be a fixed orthonormal base in $\mathbb{C}^d$.
The Hamiltonian of the system is assumed to have the form
\begin{equation}
H\;=\; \sum_{i=1}^{d} \epsilon_i |i\rangle\langle i| \otimes 1_\mathcal{F} + 1_d \otimes H_f +V \label{ham}
\end{equation}
with
\begin{equation}
H_f\;=\; \int dk \omega(k) a^*(k) a(k) \label{hamfock}
\end{equation}
while $V$ is the interaction which is typically chosen in the form
\begin{equation}
V\;=\; \sum_{i,j=1}^d |i\rangle\langle j| \otimes a^*(f_{ij })+{\rm h.c.} \label{int}
\end{equation}
Here $a(k), a^*(k)$, $k\in \mathbb{R}^d$ are annihilation and creation operators
on $\mathcal{F}$ and the following notation is used
\begin{equation}
a(f)\;=\; \int dk ~ a(k) \bar f(k)\,,
\end{equation}
where
\begin{equation}
f(k)\in L^2(\mathbb{R}^d)\,.
\end{equation}
It is clear that the Schr\"odinger equation with the above Hamiltonian cannot
be solved since one has to deal with the whole Fock space.

In order to restrict the reservoir degrees of freedom the interaction Hamiltonian
has to be simplified. Typically, one goes over to the so called $V$-models \cite{14},
i.e.~one assumes that not all transitions are allowed. More precisely one assumes that
\begin{equation}
f_{ij}\ne 0 \,, \quad  i= 2,\ldots, d\,,
\end{equation}
and all the remaining form-factors vanish.
In this case there are invariant sectors in $\mathbb{C}^d\otimes \mathcal{F}$.
In each sector the corresponding Schr\"odinger equation can be solved. However,
there is a different way to simplify the interaction Hamiltonian. Let $|\Omega\rangle$
be the vacuum state of the boson field, and let
\begin{equation}
P_0 \;=\; 1_S \otimes |\Omega\rangle\langle\Omega| \label{p0}
\end{equation}
be the projection operator on $\mathbb{C}^d\otimes \mathcal{F}$.
Following \cite{22} let us introduce the approximation $\widetilde V$ of $V$ as follows
\begin{equation}
V\;\longrightarrow\; \widetilde V \;=\; P_0 V + VP_0 -P_0 VP_0\,. \label{tildeV}
\end{equation}
In the case of the interaction $V$ given in the form \eqn{int}, one finds
\begin{equation}
\widetilde V\;=\;\sum_{i,j=1}^d |i\rangle\langle j| \otimes a^*(f_{ij}) |\Omega\rangle\langle \Omega| + {\rm h.c.} \label{intham}
\end{equation}
In this case the only possible transitions in $\mathcal{F}$ due to
$\widetilde V$ are 
\be |\Omega\rangle\;\longleftrightarrow\; a^*(f_{ij}) |\Omega\rangle\,.
\label{2.10} 
\ee 
It should be pointed out that the above
approximation makes it possible to go beyond the rotating wave
approximation.

\section{Time Evolution and Reduced Dynamics}
\setcounter{equation}{0}

Let us consider a $d$-level system coupled to a bosonic reservoir
for which the interaction Hamiltonian is given by \eqn{intham}.
If the state $|\phi_t\rangle$ of the coupled system has the form
\begin{equation}
|\phi_t\rangle \;=\; \sum_{i=1}^d |i\rangle [c_i(t) |\Omega\rangle +
a^*(g_t^i) |\Omega\rangle]\,, \label{phit}
\end{equation}
the Schr\"odinger equation is reduced to a closed system of equations for $c_1(t),\ldots,c_d(t)$ 
and $g_t^1(k),\ldots,g_t^d(k)$,
\begin{eqnarray}
\dot c_k(t) &=& -i \epsilon_k c_k(t) -i \sum_{l=1}^d (f_{lk}, g_t^l) \label{cdot}
\\
\dot g_t^l &=& -i (\epsilon_l +\omega(k))g_t^l -i \sum_{n=1}^d f_{ln}(k) c_n(t)\,, \label{gdot}
\end{eqnarray}
where
\begin{equation}
(f,g)\;=\;\int dk \bar f(k) g(k)
\end{equation}
is the scalar product in $L^2(\mathbb{R}^d)$.
Moreover, the normalization condition
\begin{equation}
\langle\phi_t|\phi_t\rangle \;=\;\langle\phi_0|\phi_0\rangle=1,
\end{equation}
takes the form
\begin{equation}
\sum_{i=1}^d [|c_i(t)|^2 + (g_t^i,g_t^i)] \;=\; \sum_{i=1}^d [|c_i(0)|^2 + (g_0^i,g_0^i)]\;=\;1\,. 
\label{normcon}
\end{equation}
Equation \eqn{gdot} can also be written in the integral form
\begin{equation}
g_t^m(k) \;=\; e^{-i t (\epsilon_m+ \omega(k))} g_0^m(k) -i \int\limits_0^t ds e^{-i (t-s) (\epsilon_m+ \omega(k))}\sum_{n=0}^d f_{mn}(k) c_n(s)\,. \label{gdot2}
\end{equation}
Inserting \eqn{gdot2} into \eqn{cdot} one finds a closed system of equations for $c_1(t),\ldots$, $c_d(t)$,
\begin{equation}
\dot c_k(t) \;=\; -i \epsilon_k c_k(t) -\int\limits_0^t ds \sum_{l=1}^d M_{kl}(t-s) c_l(s) + G_k(t)\,, 
\label{cdot2}
\end{equation}
where
\begin{equation}
M_{kl}(t)\;=\;\sum_{m=1}^d\int dk \bar f _{mk} f _{ml} e^{-i t (\epsilon_m+ \omega(k))}
\end{equation}
and
\begin{equation}
G_k(t)\;=\; -i \sum_{m=1}^d \int dk \bar f _{mk}(k) g_0^m(k)  e^{-i t (\epsilon_m+ \omega(k))}.
\end{equation}
The $M_{kl}(t)$ describe the reservoir correlation functions, while $G_k(t)$ corresponds to correlations between the reservoir and the initial state. Under the assumption that the initial state is given in the form
\begin{equation}
|\phi_0\rangle \;=\; \sum_{i=1}^d |i\rangle \otimes[c_i(0) |\Omega\rangle + a^*(g_0^i)|\Omega\rangle]\label{phi0}
\end{equation}
the total state $|\phi_t\rangle$ is given by \eqn{phit}
in terms of the solution of the equations \eqn{cdot2}, \eqn{gdot2}. It follows from \eqn{phit}
that the reduced evolution is given by
\begin{equation}
{\rm Tr}_R |\phi_t\rangle\langle\phi_t| \;=\; 
\sum_{i,j=1}^d |i\rangle\langle j| [\bar c_j(t) c_i(t) + (g_t^j,g_t^i)]\,. 
\label{red}
\end{equation}
In order to find explicitly the reduced evolution it is clear that the following correlation functions
\begin{eqnarray}
a_{mn,pq}(t)&=&\int dk \bar f_{mn}(k) f_{pq}(k) e^{-i \omega(k)} \label{acor}\\
b_{mn,p}(t)&=&\int dk \bar f_{mn}(k) g_0^p(k) e^{-i \omega(k)} \label{bcor}\\
c_{p,q}(t)&=&\int dk \bar g_{0}^p(k) g_0^q(k) e^{-i \omega(k)} \label{ccor}
\end{eqnarray}
should be known.

From the above definitions of the correlation functions, it follows
that they satisfy the following positivity condition
\begin{eqnarray}
&&\hspace*{-15mm}
\sum\limits_{\alpha mn}\sum\limits_{\beta pq}a_{mn,pq}(t_{\alpha
}-t_{\beta })\bar{x}_{mn}\left( t_{\alpha }\right) x_{pq}\left( t_{\beta
}\right) +  \nonumber \\
&&\sum\limits_{\alpha r}\sum\limits_{\beta s}c_{r,s}(t_{\alpha }-t_{\beta })%
\bar{x}_{r}\left( t_{\alpha }\right) x_{s}\left( t_{\beta }\right) +
\label{3.15} \\
&&\hspace*{15mm}
\sum\limits_{\alpha mn}\sum\limits_{\beta r}b_{mn,r}(t_{\alpha }-t_{\beta })%
\bar{x}_{mn}\left( t_{\alpha }\right) x_{r}\left( t_{\beta }\right) +\mathrm{%
c.c.} \;\geq\; 0  \nonumber
\end{eqnarray}%
for all sequences of complex numbers $x_{mn}(t_{1})
,x_{mn}(t_{2}), x_{mn}(t_3) ,\ldots $ and $x_{p}(t_{1})$, 
$x_{p}(t_{2}) ,\ldots  .$

By the Bochner theorem the correlation functions (\ref{acor})--(\ref{ccor}) 
can be presented in the form%
\begin{eqnarray}
a_{mn,pq}(t) &=&\int\limits_{-\infty }^{\infty }d\omega e^{-i\omega
t}J_{mn,pq}(\omega )\,,  \label{3.17} \\
b_{mn,p}(t) &=&\int\limits_{-\infty }^{\infty }d\omega e^{-i\omega
t}J_{mn,p}(\omega )\,,  \label{3.18} \\
c_{pq}(t) &=&\int\limits_{-\infty }^{\infty }d\omega e^{-i\omega t}J_{pq}(\omega )\,.
\label{3.19}
\end{eqnarray}
The matrices $J_{mn,pq}(\omega )$, $J_{mn,p}(\omega )$, $J_{pq}(\omega )$ define the
spectral density matrix such that for all sequences $\{x_{mn}\}$, $\{x_{p}\}$
and $\omega \in \mathbb{R}$ the relation
\begin{equation} \label{3.20}
\sum\limits_{mn}\sum\limits_{pq}J_{mn,pq}(\omega )\bar{x}_{mn}x_{pq}+
\sum\limits_{rs}J_{rs}(\omega )\bar{x}_{r}x_{s}+
\sum\limits_{mnr}J_{mn,p}(\omega )\bar{x}_{mn}x_{r}+\mathrm{%
c.c.} \;\geq\; 0 
\end{equation}
holds. Thus, choosing the spectral density matrix $J(\omega )$ one
can model the time evolution of the reduced state.

One can distinguish the special type of initial conditions,
\begin{equation}
|\phi _{0}\rangle \;=\;\Big( \sum_{i=1}^{d}|i\rangle c_{i}(0)\Big) \otimes
|\Omega \rangle\,,   \label{3.21}
\end{equation}
i.e., the simple factorized ones. In this case \eqn{cdot2} have
the form
\begin{equation}
\dot{c}_{k}(t)\;=\;-i\epsilon
_{k}c_{k}(t)-\sum_{l=1}^{d}\int\limits_{0}^{t}dsM_{kl}(t-s)c_{l}(s)\,,  
\label{3.22}
\end{equation}
while the relation \eqn{gdot2} reads%
\begin{equation}
g_{t}^{m}(k)\;=\;-i\int\limits_{0}^{t}ds\sum_{n=0}^{d}f_{mn}(k)e^{-i\left( t-s\right)
(\epsilon _{m}+\omega (k))}c_{n}(s)\,.  \label{3.23}
\end{equation}
Equations $\left( \ref{3.22}\right) $ can be solved using the Laplace
transform method, and their solution can be written in the form:%
\begin{equation}
c_{k}(t)\;=\;\sum_{l=1}^{d}L_{kl}(t)c_{l}(0)\,.  \label{3.24}
\end{equation}
On the other hand, from $\left( \ref{3.23}\right) $ it follows that%
\begin{equation}
( g_{t}^{k},g_{t}^{h}) \;=\;\sum_{m,n=1}^{d}R_{km,hn}(t)\bar{c}%
_{m}(0)c_{n}(0)\,.  \label{3.25}
\end{equation}
It is worth noting that in the case considered the correlation functions $%
\left( \ref{acor}\right) $ are needed.
Using \eqn{red}, $\left( \ref{3.24}\right) $ and $\left( \ref{3.25}\right) ,$ one finds
that the reduced evolution is given in the form%
\begin{equation}
\mathrm{Tr}_{R}\left\vert \phi _{t}\right\rangle \left\langle \phi
_{t}\right\vert \;=\;\sum\limits_{jm}\sum\limits_{in}\left\vert i\rangle
\langle j\right\vert S_{jm,in}\left( t\right) \bar{c}_{m}(0)c_{n}(0)\,,
\label{3.26}
\end{equation}%
where%
\begin{equation}
S_{jm,in}\left( t\right) \;=\;\bar{L}_{jm}(t)L_{in}(t)+R_{jm,in}\left( t\right) .
\label{3.27}
\end{equation}
The relation (\ref{3.26}) can  also be rewritten in the form%
\begin{equation}
\mathrm{Tr}_{R}\left\vert \phi _{t}\right\rangle \left\langle \phi
_{t}\right\vert \;=\;\sum\limits_{ijm}A_{t}\left( \left\vert i\right\rangle
\left\langle j\right\vert \right) c_{i}(0)\bar{c}_{j}(0)\,,
\end{equation}%
where the relation
\begin{equation}
A_{t}\left( \left\vert i\right\rangle \left\langle j\right\vert \right)
\;=\;\sum\limits_{mn}\left\vert m\right\rangle \left\langle n\right\vert
S_{nj,mi}\left( t\right)
\end{equation}%
defines a map%
\begin{equation}
A_{t}:M_{d}\longrightarrow M_{d}\,,\quad t\geq 0\,,
\end{equation}%
which is completely positive and trace preserving by construction.

The discussion of time evolution under initial conditions different from $%
\left( \ref{3.21}\right) $ is, in general, very complicated and will be
presented on a simple model.

\section{Spin-Boson Model beyond the Rotating Wave Approximation}
\setcounter{equation}{0}
The model is specified by the Hamiltonian
\be
H=\omega\sigma^+\sigma^- + \int dk \omega(k) a^*(k) a(k) + \sigma^-
\otimes a^*(f) |\Omega\rangle\langle\Omega| + \sigma^+ \otimes
a^*(h)|\Omega\rangle\langle\Omega| + \rm{h.c.} \label{4.1}
\ee
on $\mathbb{C}^2 \otimes \mathcal{F}$ and can be obtained from the
spin-boson Hamiltonian containing anti-resonance interaction using
the method presented in Sect.~3. The wave function can be chosen
in the form
\be
|\phi_t\rangle \;=\; (c_0(t)|0\rangle + c_1(t) |1\rangle)\otimes
|\Omega\rangle +|0\rangle\otimes a^*(g_t^0) |\Omega\rangle +
|1\rangle) \otimes a^*(g_t^1) |\Omega\rangle \label{4.2}
\ee
since the interaction in \eqn{4.1} allows only the transition
between the vacuum state and 1-boson states. The Schr\"odinger
equation is equivalent to the following set of equations for
$c_0(t)$, $c_1(t)$ and $g_t^0(k)$, $g_t^1(k)$:
\beqa
\dot c_0&=&-(h, g_t^1)\,, \label{4.3} \\
\dot c_1&=& -i \omega c_1(t) -i (f, g_t^0)\,, \label{4.4} \\
\dot g_t^0(k)&=& -i \omega (k) g_t^0(k) -i f(k) c_1(t))\,, \label{4.5} \\
\dot g_t^1(k)&=& -i(\omega+ \omega (k)) g_t^1(k) +h(k)  c_0(t))\,.
\label{4.6}
\eeqa
Equations \eqn{4.4} and \eqn{4.5} are the same as in the case of
spin-boson model in the rotating wave approximation. On the other
hand \eqn{4.3}, \eqn{4.6} describe the effect of anti-resonance
interaction. It follows from \eqn{4.3}--\eqn{4.6} that our system has
two constants of motion:
\be
|c_0(t)|^2 + (g_t^1,g_t^1) \;=\;|c_0(0)|^2 + (g_0^1,g_0^1)\;=\; p_0\;\ge\;
0\label{4.7}
\ee
and
\be
|c_1(t)|^2 + (g_t^0,g_t^0) \;=\;|c_1(0)|^2 + (g_0^0,g_0^0)\;=\; p_1 \;\ge\;
0\label{4.8}
\ee
Moreover, the normalization condition
\be
\langle\phi_t|\phi_t\rangle\;=\;\langle\phi_0|\phi_0\rangle\;=\;1\,,
\label{4.9}
\ee
takes the form
\be
p_0+p_1\;=\; 1\,. \label{4.10}
\ee
The reduced state ${\rm Tr}_R \langle\phi_t|\phi_t\rangle$ can be
written as follows
\be
\rho(t) \;=\;{\rm Tr}_R \langle\phi_t|\phi_t\rangle\;=\; \sum_{i,j= 0}^1 \langle
i|j\rangle\rho_{ij}(t)\,, \label{4.11}
\ee
where
\beqa
\rho_{00}(t) &=& |c_0(t)|^2 + (g_t^0,g_t^0) \label{4.12} \\
\rho_{11}(t) &=& |c_1(t)|^2 + (g_t^1,g_t^1) \label{4.13} \\
\rho_{01}(t) &=&\bar \rho_{10}(t) \;=\; c_0(t) \bar c_1(t) +
(g_t^1,g_t^0)\,. \label{4.14}
\eeqa
Using \eqn{4.7} and \eqn{4.8} the diagonal elements of $\rho(t)$ can
be written in the form
\beqa
\rho_{00}(t)
&=&|c_0(t)|^2+|c_1(0)|^2-|c_1(t)|^2+(g_0^0,g_0^0)\label{4.15} \\
\rho_{11}(t)&=&1-\rho_{00}(t)\,.\label{4.16}
\eeqa
It should be pointed out that the equations \eqn{4.3}, \eqn{4.6} and
\eqn{4.4}, \eqn{4.5} are independent. However, there exists coupling
in terms of correlation functions which are needed to calculate the
reduced evolution. Eliminating $g_t^0(k) $ and $g_t^1(k) $ from
\eqn{4.5} and \eqn{4.6} the following equations for $c_0(t)$ and
$c_1(t)$ can be derived
\beqa
\dot c_0(t)&=&\int\limits_0^t ds~ m_0(t-s) c_0(s) + n_0(t)\,, \label{4.17}\\
\dot c_1(t)&=& -i\omega c_1(t) -\int\limits_0^t ds~ m_1(t-s) c_1(s) +
n_1(t)\,,\label{4.18}
\eeqa
where
\beqa
m_0(t)&=& \int dk ~|h(k)|^2 e^{-i t (\omega+\omega(k))}, \label{4.19}\\
m_1(t)&=& \int dk ~|f(k)|^2 e^{-i t \omega(k)}\label{4.20}
\eeqa
and
\beqa
n_0(t)&=& -i \int dk~ \bar h(k) g_0^1(k)  e^{-i t (\omega+\omega(k))}\,, \label{4.21}\\
n_1(t)&=& -i \int dk~ \bar f(k)g_0^0(k) e^{-i t
\omega(k)}\label{4.22}\,. 
\eeqa 
In order to find out an explicit form
of $\rho(t)$ a detailed knowledge of all the correlation functions
is required. However, general properties of $\rho_{00}(t)$ and
$\rho_{11}(t)$ can be deduced in the following manner. Suppose
that $A_0(t)c_0(0)$ and $A_1(t)c_1(0)$ are general solutions of
homogeneous equations \eqn{4.17} and \eqn{4.18}, respectively. Then one
has 
\be 
c_\alpha(t)\;=\;A_\alpha(t)c_\alpha(0) +\int\limits_0^t ds ~
A_\alpha(t-s)n_\alpha (s) \label{4.23} 
\ee 
with $\alpha=0,1$ and
\be 
\rho_{00}(t)\;=\; |c_0(t)|^2 + |c_1(0)|^2-|c_1(t)|^2 +
(g_0^0,g_0^0)\,. \label{4.245} 
\ee 
It follows from \eqn{4.23} that
the time dependence of $\rho_{00}(t)$ is related to the initial
conditions. Moreover, the asymptotic behaviour of $\rho_{00}(t)$
can be easily found as 
\be 
\lim_{t\rightarrow \infty}
\rho_{00}(t)\;=\; |c_1(0)|^2 + (g_0^0,g_0^0)\label{4.25} 
\ee 
since,
typically, one has 
\be 
\lim_{t\rightarrow \infty}c_\alpha(t)
=0\,, \quad  \alpha=0,1\,. \label{4.26} 
\ee 
The formula \eqn{4.25}
shows that the asymptotic state is determined by the initial
conditions, i.e., the equilibrium state does not exist. In the
case $g_0^0(k)=g_0^1(k)=0$ the time evolution is given in terms of
a completely positive and trace preserving but not relaxing map.

The spin-boson model is recovered if one puts
$h(k)=0=g_0^1(k)=g_t^1(t)$. In this case we have
\be
|c_0(t)|^2=|c_1(0)|^2
\ee
and, using the normalization condition \eqn{4.8} one finds
\begin{eqnarray}
\rho_{00}(t)&=&1-|c_1(t)|^2\\
\rho_{11}(t)&=&|c_1(t)|^2\,.
\end{eqnarray}

\section*{Acknowledgments}

A.\,K. thanks Beppe Marmo for his warm hspitality and the MECENAS project for financial support during his stay in Dipartimento di Scienze Fisiche dell'Univer\-sit\`a `Federico II' in Napoli.

\end{document}